\documentclass[journal]{IEEEtran}
\ifCLASSINFOpdf
\else
\fi
\usepackage{makecell}
\usepackage{bbding}
\usepackage{graphicx}
\usepackage{amsmath}
\usepackage{amssymb}
\usepackage{float}
\usepackage{stfloats}
\usepackage{algorithm}
\usepackage{algpseudocodex}
\usepackage{cite} 
\usepackage{bm}
\usepackage{soul}
\usepackage{color}
\usepackage{upgreek}
\usepackage{multirow}
\usepackage{multicol}
\usepackage{siunitx}
\def\undb#1{\mbox{\bf{#1}}}
\begin{document}
\bstctlcite{bstctl:nodash}
%
\title{Spike Talk: Genesis and Neural Coding Scheme Translations}
%
%
%

\author{ 
        Subham Sahoo,~\IEEEmembership{Senior Member, IEEE}
\thanks{Subham Sahoo is an Associate Professor with the Department of Energy, Aalborg University, Denmark. (e-mail: sssa@energy.aau.dk)
}}
\maketitle

%
\begin{abstract}
Although digitalization of future power grids offer several coordination incentives, the reliability and security of information and communication technologies (ICT) hinders its overall performance. In this paper, we introduce a novel architecture \texttt{{Spike Talk}} via a unified representation of power and information as a means of data normalization using spikes for coordinated control of microgrids. This grid-edge technology allows each distributed energy resource (DER) to execute decentralized secondary control philosophy independently by interacting among each other using power flow along the tie-lines. Inspired from the field of computational neuroscience, \texttt{\textbf{Spike Talk}} basically builds on a fine-grained parallelism on the information transfer theory in our brains, particularly when neurons (modeled as DERs) transmit information (inferred from power streams measurable at each DER) through synapses (modeled as tie-lines). Not only does \texttt{\textbf{Spike Talk}} simplify and address the current bottlenecks of the cyber-physical architectural operation by dismissing the ICT layer, it provides intrinsic operational and cost-effective opportunities in terms of infrastructure development, computations and modeling. Hence, this paper provides a pedagogic illustration of the key concepts and design theories. Since we focus on coordinated control of microgrids in this paper, the signaling accuracy and system performance is studied for several neural coding schemes responsible for converting the real-valued local measurements into spikes. 

\end{abstract}

\begin{IEEEkeywords}
Neuromorphic computing, microgrids, hierarchical control, spiking neural network.
\end{IEEEkeywords}

%
\IEEEpeerreviewmaketitle

\section{Introduction}
\begin{figure}[!t]\centering
	\includegraphics[width=\linewidth]{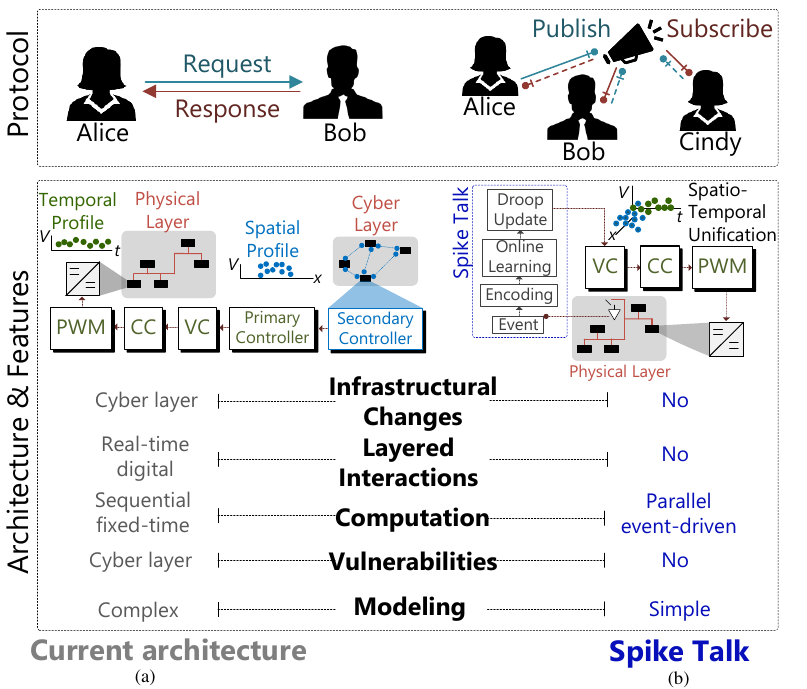}
	\caption{As opposed to the current cyber-physical architecture in (a), an innovative co-transfer technology (b) \texttt{\textbf{Spike Talk}} emanates out of novel protocol, namely \textit{publish-subscribe}, that supports the inferential exchange of information from power flows in the network. It is a software-defined methodology that offers cognitive abilities to DERs and their coordination by providing numerous benefits without requiring any infrastructural changes.}
    
    \label{FIG_422}
\end{figure}

%
%
%
%
\IEEEPARstart
{D}{igitalization} of power electronic dominated grids has brought new data-driven directions in health monitoring and coordination of ever-scaling distributed energy resources (DERs). As we move towards decentralized generation, the control and reliability of DERs demand orchestration from information and communication technologies (ICT) to assist with system coordination, as shown in Fig. \ref{FIG_422}(c). However, these digitalization efforts not only have increased cyber attack vulnerabilities \cite{8901166}, but have added more unreliability metrics in the form of latency, data dropouts, etc. \cite{espina2023consensus}. This mandates an intrinsic coordination principle for power grids that can effectively unify power \& information to enhance system reliability.\par

Apart from the communication reliability issues, the layered hierarchy of the ICT layer overlaying the power grids fabricates strong dependencies on each other, thereby introducing modeling complexity \cite{platzer}. With increased penetration of ultra-efficient power electronic converters operating at a switching frequency within 100-500 kHz enabling faster control response, it simultaneously mandates upgradation of the ICT infrastructure to level up the coordination efforts. With modeling complexities of the cyber-physical interactions and infrastructural changes still in place as key research questions to be solved, co-transfer technologies that can modulate data over power streams emerged as a monolithic solution to all the unreliability issues in the cyber layer above \cite{petar1}, \cite{hutzell}. The basic principle behind co-transfer technologies is to exploit each DER\footnote{Since power electronic converters are equipped in each DER, we use the term DER to represent DC/DC power electronic converter throughout this paper.} with a power electronic interface as a transceiver and the tie-line between them as a communication channel. Finally, control parameter adaption is performed at each DER with physical disturbances set as the triggering mechanism. 

Differently from the conventional power line communication\cite{plc}, \textit{Power Talk} proposed in \cite{angjelichinoski2016multiuser}, allows to send information by modulating the bus voltages and corresponding droop parameter and consequently create power deviations, that can be detected and decoded by other converters. This one-way communication model is then extended in \cite{plc2} for its constellation design and error analysis in the signaling and detection space. From a physical layer standpoint, this technology demystifies the re-design of the primary controller into a communication paradigm with signal transceiver abilities. Another emerging conversion technology, namely \textit{Talkative Power} \cite{tpc1,tpc3} embeds information into the pulse width modulation (PWM) stage instead, with a clear rationale of extracting information as harmonics on eliminating the filtering stages. Not only the communication rate can be improved by modifying the switching characteristics, but it does not require additional hardware. An overview behind various co-transfer technologies that transmit power \& data along tie-lines can be found in \cite{tpc2}. Although they guarantee good reliability, they are highly dependent on graph theoretic concepts for ensuring low error rates and convergence of information transfer. Furthermore, they follow the \textit{request-respond} two-way communication protocol (as shown in Fig. \ref{FIG_422}), which serves as a common push-pull communication approach and limits co-transfer scalability among multiple DERs in large power electronic systems. 

From a cyber-physical system perspective, unified transmission of power \& information along the physical layer itself in Fig. \ref{FIG_422} is capable of solving many issues altogether with: (i) no infrastructural changes, (ii) dismissal of the cyber layer, its inconsistent performance, vulnerabilities and interactions with the physical layer, (iii) simplified modeling. With the \textit{request-response} protocol limiting the scalability of information transfer in large systems, we firstly explore the \textit{publish-subscribe} communication protocol \cite{petar} for power grids in Fig. \ref{FIG_422} that facilitates a \textit{fluctuation-driven} regime, such that the control parameters adaptation using co-transfer technologies are governed solely based on the proximity of selective DERs to the disturbance's spatial location and impact itself. The protocols in Fig. \ref{FIG_422} can be explained clearly using the simple example of a newsletter, wherein although Alice, Bob, and Cindy are subscribed to it, they read it only when it interests/concerns any of them. Following up on our preliminary evidence to locally estimate remote signals in power grids using task-aware semantic metrics \cite{kirti,kirti2}, we can integrate the \textit{publish-subscribe} based inferential protocol for the same task, only when there is an event. Although the abovementioned work was primarily designed to defend against cyber anomalies in a cyber-physical system setup in Fig. \ref{FIG_422}, the root cause behind the problem, i.e., communication traffic, jamming, etc. still persists due to the presence of the cyber layer itself. As we transit towards the future power grid architecture as a single-layer architecture in Fig. \ref{FIG_422} enabling independence from the cyber layer, we need an effective tool at each node that can locally synthesize and adapt a decentralized secondary control policy from the power flow data itself. 

In this paper, we draw inspiration from the field of computational neuroscience in understanding how several independent sensors act together in our brain to understand perception. This can be attributed to a comprehensive information mechanism that accounts for spatio-temporal data and its processing in our brain. Using a similar operational philosophy without being circumscribed by graph theory, we propose \textbf{\texttt{Spike Talk}}, which treats each DER as an event-driven sensor to infer global information and adapt their local controllers accordingly only using power flows as the medium between them. It is worth notifying that \textbf{\texttt{Spike Talk}} fundamentally differs from \textit{Power Talk} and \textit{Talkative Power} due to its new \textit{publish-subscribe} protocol described above. As a result, the proposed co-transfer technology does not demand transfer of information from other nodes, but should rather be seen as an on-top software-defined control logic that extracts task-oriented data from power flows and its dynamics to adapt DERs' response in the network. More details on a comprehensive comparative analysis between these co-transfer technologies can be obtained in our recent work \cite{mag,moore,infer}. Unlike our previous work \cite{neuro} in showcasing its energy-efficient computational abilities and online learning, we deep dive into the following aspects in this paper:
\begin{enumerate}
    \item \textit{\textbf{Genesis:}} Since co-transfer of power and information using Spike Talk is fundamentally driven by the information transfer mechanism between neurons in our brains, we first provide a pedagogic illustration to point out the electrical modeling similarities between neurons and DERs. We then use the component-level modeling preliminaries and extend them to a system-level analogy for both neuronal networks and power grids. Firstly, the electrical model of leaky integrate-and-fire (LIF) neuron and a generalized equivalence with converter topology is explored for structural mapping. Secondly, a selective process of \textit{task-oriented} sparse data is collected for online training, where we leverage the $RC$ dynamics of the LIF neuron and translate the same data collection philosophy for each DER. Using Spike Talk, the entire power electronic grid can be emulated as a multi-sensory perceptive network with DERs at each node inferring via power flows only.
    \item \textit{\textbf{Neural Coding Schemes:}} Using the distributed computation facility established at each DER by Spike Talk, the system is adapted only during events represented by sparse dynamic data. Hence, the most critical stage is encoding real analog-valued measurements into sparse binary spikes. This stage affects the following aspects simultaneously:
    \begin{itemize}
        \item signalling accuracy and system performance,
        \item computation time (sparsity in data),
        \item computation overheads (online learning).
    \end{itemize}
\end{enumerate}
{Given that the secondary controller response is typically expected in seconds \cite{sahoo2017distributed}, Spike Talk is well equipped for this requirement, since its online learning capability can be achieved within general neuronal reaction time of 100-200 ms.}

In this paper, we validate these two aspects in an exemplary DC microgrid. With the comparative evaluation of Spike Talk and the existing co-existing technologies already covered in \cite{neuro}, we particularly emphasize on its modeling part and signalling consequences due to different neural coding schemes in this paper. With the modeling preliminaries as a key foundation, other stages of Spike Talk will be introduced later in future scope of this article. 
\begin{figure*}[t]\centering
	\includegraphics[width=0.95\linewidth]{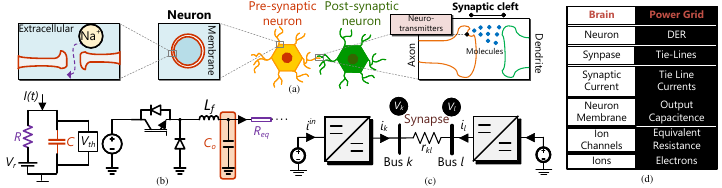}
	\caption{Modeling analogies between neurons and DER: (a) neuron membrane and synpases, (b) the electrical modeling of the membrane as a $\mathrm{RC}$ circuit, which is retrofit into the output capacitor and equivalent tie-line resistance seen from a DC/DC converter, (c) Similar to the pre-synpatic and post-synaptic connection in (a), two DC DERs (emulated as neurons) are connected to each other via tie-lines (emulated as synpase) to transmit ingrained information via power, (d) structural mapping of each component in the brain and power grid to construct a duality principle.}
    \label{FIG_n}
    \vspace{-6 pt}
\end{figure*}
\section{Unraveling the Modeling Analogies}
While there are a range of bio-physically accurate neuron models in the field of computational neuroscience from the most complicated ones being Hodgkin-Huxley and Izhikevich models to simple artificial neurons \cite{pooi}, we select the leaky integrate-and-fire (LIF) model since it provides a good trade-off between the biological accuracy and complexity \cite{gerster}. Unlike artificial neural networks (ANNs) process the weighted sum of inputs using an activation function, LIF neurons on the other hand integrate the weighted sum input over time with a leakage, much like a RC circuit. In translation, we exploit the same RC dynamics in output of each DER to collect sparse data and incorporate online learning. Before continuing with the sparsity of data, let us understand the modeling similarities between a LIF neuron and DERs in the next subsection.
\subsection{Component Level Modeling}
Our brain has around billions of neurons. Using multiple learning instances in the past, it can sense, perceive and then take dynamic, complex decisions quickly. For a given stimuli, neurons fire to transmit an information using their network to the brain. As shown in Fig. \ref{FIG_n}(a), two neurons are connected to each other via a synpase, which is responsible for transferring electro-chemical ions as information via their neuro-transmitters. Since the neurons are placed on either side of the synpase, they are called \textit{pre-} and \textit{post-synpatic neurons}. Furthermore, each neuron is surrounded by a thin bi-layer membrane that insulates the intracellular portion from the extracellular medium. Due to this property where two conductive solutions are separated by an insulator, this membrane can be electrically interpreted as a capacitor $\mathrm{C}$. This membrane also exhibit an impermeable property that limits the passage of ions (Na$^+$ in Fig. \ref{FIG_n}(a)) from the extracellular medium inside the neuron. However, there are specific membrane channels that are triggered to open when a current is injected into the neuron, which mimics the electric property of a resistor $\mathrm{R}$. Considering an arbitrary time-varying current $\mathrm{I(t)}$ in Fig. \ref{FIG_n}(b) injected into a neuron, it can be written as:
\begin{equation}
    \mathrm{I(t) = \frac{V_{mem}(t)}{R} + C\frac{d V_{mem}}{dt}}
\end{equation}
with $\tau=\mathrm{RC}$ being the circuit time constant, $\mathrm{V_{mem}}$ being the voltage across the membrane. Hence, the LIF neuron membrane can be modeled as a $\mathrm{RC}$ circuit, where the first and second term in (1) corresponds to the leakage part and integrate-and-fire part, respectively. In this way, the passive membrane of neuron can be described by a linear differential equation in (1). Considering the derivative $\mathrm{\frac{dV_{mem}}{dt}}$ substituted into (1) without considering the limit $\mathrm{\Delta t} \rightarrow$ 0, we get:
\begin{eqnarray}
    \mathrm{\tau \frac{V_{mem}(t + \Delta t) - V_{mem}(t)}{\Delta t} = - V_{mem}(t) + I(t)R}
\end{eqnarray}
Isolating the membrane potential for the next time-step in LHS and assuming input current $\mathrm{I(t)}$ = 0 to single out the resistive leaky dynamics, we simplify (1) into:
\begin{eqnarray}
    \mathrm{V_{mem} (t + \Delta t) = (1 - \frac{\Delta t}{\tau})V_{mem}(t)}
\end{eqnarray}
Considering the ratio of the subsequent membrane potential values, $\mathrm{V_{mem}(t + \Delta t)}$/$\mathrm{V_{mem}(t)}$ to be the decay rate $\beta$, which can then be mathematically denoted by $\mathrm{(1 - \Delta t/\tau)}$. However, it is preferred to calculate $\mathrm{\beta}$ using $\mathrm{e^{-\Delta t/\tau}}$, since $\mathrm{\Delta t << \tau}$ \cite{abbott}.

\textit{\textbf{Remark I:} Similar to the electrical $\mathrm{RC}$ circuit model of a LIF neuron in (1), we exploit the same network elements in a DC DER in Fig. \ref{FIG_n}(b) using its output capacitance $\mathrm{C_o}$ and the equivalent network resistance $\mathrm{R_{eq}}$ to formulate a physics-informed threshold based criteria for sparse data collection. The rationale behind using $\mathrm{R_{eq}}$ is that it can be adapted/controlled in alignment with power flow dynamics.} 
\subsection{System Level Modeling}
Using the component modeling for each DER in Remark I as a pre-requisite groundwork, we will now expand it to a system-level control policy in this subsection. For the current carried by the ions in Fig. \ref{FIG_n}(a), the most critical dynamic variable is the potential difference between the pre-synaptic $\mathrm{k}$ and post-synaptic neuron $\mathrm{l}$. This can be attributed to the dynamics of the potential using Kirchhoff's current law:
\begin{equation}
    \mathrm{g_{kl} (V_k - V_l)} = \underbrace{\mathrm{I_k - C_k \frac{dV_k}{dt}}}_{\text{Neuron $\mathrm{k}$}} + \underbrace{\mathrm{C_l \frac{dV_l}{dt} - I_l}}_{\text{Neuron $\mathrm{l}$}}
\end{equation}
where, the term in LHS represent the potential difference between the neuron membranes $\mathrm{k}$ and $\mathrm{l}$, whereas the terms in RHS correspond to the capacitive properties of both neurons as well as their synaptic currents $\mathrm{I_k}$ and $\mathrm{I_l}$. It is worth notifying that the $\mathrm{g_{kl}}$ in (2) denote the synaptic conductance. Using (1), (2) can be re-written for each neuron as:
\begin{equation}
        \mathrm{g_{kl} (V_k - V_l) = g_k(V^k_{mem} - V_{th}) - g_l(V^l_{mem} - V_{th})}
\end{equation}
where, $\mathrm{g_k}$ and $\mathrm{g_l}$ denote the membrane conductance of neuron $\mathrm{k}$ and $\mathrm{l}$, respectively. Using the primary controller modeling preliminaries \cite{dcm}, we can easily translate the electrical equivalent circuit of each neuron in Fig. \ref{FIG_n}(a) to that of a DC DER. Comparing the multi-neuron and multi-DER system in Fig. \ref{FIG_n}(a) and \ref{FIG_n}(c) for a system-level analogy, a duality principle can be inferred, that has been elaborated in Fig. \ref{FIG_n}(d). 

\textit{\textbf{Remark II:} With two neurons having their corresponding membrane potential across their surfaces, this can be equivalenced to the voltages across each DER. Furthermore, the synaptic cleft connects the neurons for transmission of information between them, which can be electrically interpreted as synaptic currents. Using the structural mapping in Fig. \ref{FIG_n}(d), we hypothesize to transmit ingrained information in the current flowing between two DERs by observing their dynamics and response to any disturbances.}

\begin{figure}[!thb]\centering
	\includegraphics[width=0.97\linewidth]{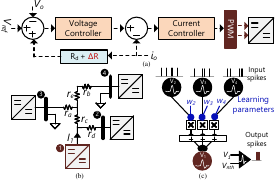}
	\caption{(a) Primary control of a DC DER with $\mathrm{R_d}$ and $\mathrm{\Delta R}$ as the static and adaptive droop values, respectively, (b) Single line diagram of the 4-bus DER based microgrid, (c) Spiking feedforward network with LIF neurons.}
    \label{FIG_DER}
\end{figure}
Using the structural mapping formalism in Remark II, the current flowing through the line with a tie-line resistance $\mathrm{r_{kl}}$ from DER $\mathrm{k}$ to DER $\mathrm{l}$ operating at a global voltage reference $\mathrm{V_{ref}}$ can be given by:
\begin{equation}
\mathrm{r^{-1}_{kl}(V_k - V_l)} = \underbrace{\mathrm{\frac{V_k - V_{ref}}{R^k_d + \Delta R_k}}}_{\text{DER $\mathrm{k}$}} - \underbrace{\mathrm{\frac{V_l - V_{ref}}{R^l_d + \Delta R_l}}}_{\text{DER $\mathrm{l}$}}
\end{equation}
where, $\mathrm{R_d}$ is a droop parameter for each DER, that can be calculated by $\mathrm{\frac{\Delta V}{I_{max}}}$ with $\mathrm{\Delta V}$ and $\mathrm{I_{max}}$ denoting the maximum allowed voltage deviation and rated current of the DER, respectively. It is worth notifying that we exploited the primary control model in Fig. \ref{FIG_DER} to simplify (4). Since $\mathrm{R_d}$ is a static parameter, our objective is to update $\mathrm{\Delta R}$ to instill a decentralized secondary control policy for each DER, which is corrected in accordance with the power flow dynamics defined as an \textit{event}. These dynamics are firstly converted into \textit{task-oriented} spikes, which are then processed by the LIF based neuronal computational unit at each DER to compute $\mathrm{\Delta R}$.

\textit{\textbf{Remark III:} Explicit network level computational burden with ANN at each node is significantly descended to a neuron level computation using the LIF neuron model for each DER, that not only decreases the computational overhead of the entire system, but also pans out as an energy-efficient and scalable data-driven control methodology for large power electronic systems.} 
\subsection{Sparse Data Collection}
The data \& computational sparsity in Spike Talk can totally be attributed to the binary nature of the data and physics-informed data collection mechanism. This is achieved simultaneously by collecting spikes that are encoded from real-valued measurements only when a dynamic threshold is reached. Spikes can be simply defined as electrical pulses, that can have multi-modal representations. For example, a single spike can simultaneously represent the voltage and current information. Based on the neuron firing criteria, a cluster of such spikes can serve to reconstruct dynamic profiles of voltages and currents. However, the said criteria can be set up such that spikes are generated in an asynchronous manner. This criteria can be determined analytically as a solution to the LIF neuron in (1).

Applying a step current input in $\mathrm{I(t)}$ that switches on at $\mathrm{t}$ = $\mathrm{t_0}$, the general solution can be denoted as:
\begin{equation}
    \mathrm{V_{mem}(t) = I(t)R + [V_0 - I(t)R]e^{-t/\tau}}
\end{equation}
where, $\mathrm{V_0}$ is the steady-state voltage. For $\mathrm{V_{mem}}$ = 0 V, we can re-write (7) as:
\begin{equation}
    \mathrm{V_{mem} (t) = I(t) R[1-e^{-t/\tau}]}
\end{equation}
Replacing $\mathrm{e^{-\Delta t/\tau}}$ with $\beta$ in (8) as per (3), we can exploit a non-physiological assumption to convert the term (1-$\beta$) into a learnable weight matrix $\undb{w}$ to obtain:
\begin{eqnarray}
    \mathrm{\undb{w} \undb{V}_{mem}(t) = \undb{I}(t)}
\end{eqnarray}
where, $\mathrm{\undb{V}_{mem}}$ and $\mathrm{\undb{I}}$ denote the vector representations of membrane potential and currents from each neuron. As a result, the membrane voltage difference scaled by the synaptic conductances in \undb{w} constitute the input current \undb{I}, which has been shown in the spiking feedforward network in Fig. \ref{FIG_DER}(c). Given that (9) commands good biological plausibility and accuracy, it is also highly congruent to the networked DC power flow theory in a multi-DER microgrid in Fig. \ref{FIG_DER}(b), such that the potential difference between each DER buses constitute the current flow between them. It is worth notifying that (9) is highly congruent to the DC-OPF equation $\mathrm{\undb{B}{\theta} = \undb{P}}$ \cite{bergen}, where the difference in bus angles $\mathrm{\theta}$ leads to power flow between them.

With (9) serving as an instantaneous representation of the synaptic current for a given spike, we can analyze the membrane potential of the neighboring DER (post-synaptic neuron) with synaptic current as the input using:
\begin{eqnarray}
   \mathrm{ \undb{V}_{mem}(t+1)} = \underbrace{\mathrm{\beta \undb{V}_{mem}(t)}}_{\text{Decay}} + \underbrace{\mathrm{\undb{w}\undb{X}(t)}}_{\text{Input}} - \notag \\ \underbrace{\mathrm{S_{out}(\beta \undb{V}_{mem}(t) + \undb{w} \undb{X}(t))}}_{\text{Reset-to-zero}} 
\end{eqnarray}
\begin{figure}[!thb]\centering
	\includegraphics[width=\linewidth]{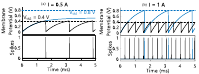}
	\caption{LIF neuron model for generating spikes: (a) membrane potential for a synpatic current of 0.5 A only generates spikes with the lower voltage threshold $\mathrm{V_{th1}}$ = 0.4 V, (b) when the synaptic current is increased to 1 A, regularly interspaced spikes are generated for both the voltage thresholds.}
    \label{FIG_42}
\end{figure}
where, \undb{X}(t) is an unweighted current from remote DERs. Considering different input current step inputs in (10), it can be seen in Fig. \ref{FIG_42}(a) that the membrane potential rises upto a resting potential $\mathrm{V_{th2}}$ = 0.4 V, which is where the reset mechanism brings it back to zero potential, thereby generating output spikes. In this paper, we have simply used reset-to-zero method as the reset mechanism. More details on reset mechanisms with LIF neuron firing can be found in \cite{pooi}. However for the same input current of 0.5 A introduced as a step input with $\mathrm{V_{th1}}$ raised to 0.8 V, the membrane potential exponentially relaxes to $\mathrm{I(t)R}$, which is smaller than $\mathrm{V_{th1}}$ leading to no output spikes. At the same time, when the input current is increased to 1 A in Fig. \ref{FIG_42}(b), spikes are generated at different spacing intervals for the resting potential $\mathrm{V_{th1}}$ = 0.8 V. Hence, as opposed to continuously sampled real membrane potential values in Fig. \ref{FIG_42}, we convert them into periodic interspaced spikes using a neuron firing criteria. Since it returns periodic spikes even for a constant value of input current, it still doesn't sufficiently enforce an \textit{event-driven} sparsity. The sparsity in spikes can be further enhanced by using an asynchronous neuron firing criteria, commonly known as \textit{neural coding schemes}, that will be described in detail in the next section.
\section{Neural Coding Schemes and Deployment}
In this section, we learn about encoding of real-valued signals into binary spikes, which is a common way of data representation in our brain and its perception. These perceptive signatures are basically a statistical reflection of either the signal magnitude or latency, which needs encoding laws that can process big data into few and qualitative batches of spikes.
\begin{figure}[!thb]\centering
	\includegraphics[width=0.75\linewidth]{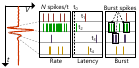}
	\caption{Neural coding schemes to represent dynamics in time-domain signals using: (a) rate coding, (b) latency coding, (c) burst coding. It is important to note that burst is a group of short inter-spaced interval (ISI) spikes.}
    \label{FIG_code}
\end{figure}
\subsection{Rate Coding}
With extensive studies already carried out determining the rate-coding corresponding to the pixel intensity in an image, we conceptualize a similar analogy to time-domain signals based on the magnitude. Since our focus in this paper is on inferring remote information only during events, we can deduce the generation of rate-coded spikes $\Omega_r$ using:
\begin{eqnarray}
  \mathrm{  \Omega_r} = \begin{cases}
        1 \ \textbf{if} \ \mathrm{\dot{V}_{k}(t)} \geq \mathrm{\kappa_1 e_v(t)} \\
        0 \ \textbf{else}
    \end{cases}
\end{eqnarray}
where, $\mathrm{e_v(t)}$ = [$\mathrm{V_{ref}}$ - $\mathrm{V_k(t) + I(t)(R_d + \Delta R(t))}$] is the voltage controller error in Fig. \ref{FIG_DER}(a) and $\kappa_1$ is a tuning parameter. As shown in Fig. \ref{FIG_code}(a), the spike train profile is different with respect to the different dynamic stages. This coding scheme is quite straight-forward, since (11) doesn't necessarily instill any dependency on model parameters.
\subsection{Latency Coding}
Latency coding is derived from a notion of elapsed time to indirectly denote the spike timing based on the feature intensity in a signal. For example, in Fig. \ref{FIG_code}(b), although only one spike is generated for each time window, it is the elapsed time from the initial set-point $t_0$ for each spike, that becomes an indirect representative of the signal magnitude. This is a very widely used signal coding technique to process sparse data for spiking neural networks (SNNs) \cite{blake}. For a constant current injection in (8), $V_{mem}$ will exponentially reach a steady-state value of $\mathrm{I(t)R}$. Given that a spike is emitted when (8) reaches a threshold of $V_{th}$, we can then obtain $t$ using:
\begin{eqnarray}
   \mathrm{ t = \tau [\text{ln}(\frac{IR}{IR - V_{th}})]}.
\end{eqnarray}
Using (12), it can be inferred that the larger input current, the convergence of $\mathrm{V_{mem}(t)}$ into $\mathrm{V_{th}}$ will be faster with fleeting spikes.

\textit{\textbf{Remark IV:} Since (11) is dependent on a nodal variable voltage that varies for each DER, rate coding can generally depict nodal voltage dynamics, whereas the spike generation using latency coding in (12) is primarily dependent on currents flowing between two DERs, it is more appropriate for network level depiction.}

Hence, the spikes generation criteria using latency coding with the steady-state feature $\mathrm{IR}$ can simply be given by:
\begin{equation}
    \mathrm{t(IR)} = \begin{cases}
        \mathrm{\tau [\text{ln}(\frac{IR}{IR-V_{th}})], \ \textbf{if} \ IR > V_{th}} \\
        \infty,  \ \textbf{else}
    \end{cases}
\end{equation}
\subsection{Burst Coding}
\begin{figure*}[!thb]\centering
	\includegraphics[width=0.85\linewidth]{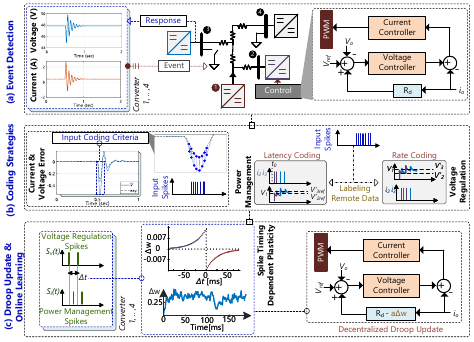}
	\caption{Implementation overview of \texttt{\textbf{Spike Talk}} -- (a) event detection: each DER will have its local coding mechanism corresponding to each event. Since there is no cyber layer, the controllers are only equipped with primary control scheme, (b) the input spikes are continuously generated for every dynamics, however, the \textit{semantic} output spikes are filtered using different network based voltage thresholds. These spikes are finally processed in tandem with different control objectives, (c) Continual online learning for each DER -- the nodal currents and voltages are firstly being processed into spikes by rate and latency coding based on Remark IV. Finally, the relative timing based plasticity rules \cite{neuro} were used to adapt the adaptive droop parameter $\mathrm{\Delta R} = \mathrm{a \Delta w}$. It is worthy notifying that the update of $\mathrm{\Delta R}$ is a system-centered approach, which is administered by the power dynamics seen by each DER.}
    \label{FIG_12}
\end{figure*}

With several bottlenecks in efficiency behind information transmission, burst coding has been an effective tool for fast and energy-efficient inferences using high frequency burst spikes. As shown in Fig. \ref{FIG_code}(c), burst spikes are a group of low interspaced interval (ISI) spikes next to each other. Basically, burst coding can simply be executed by converting the constant tuning parameter $\mathrm{\kappa}$ in (11) into a burst function-based parameter $\mathrm{\kappa_b}$ to obtain burst spikes $\mathrm{\Omega_b}$ using:
\begin{equation}
    \mathrm{\kappa_b} (t) = \begin{cases}
       \mathrm{ \Xi \kappa_b (t-1)}, \ \textbf{if} \ \mathrm{\Omega_b(t-1)} = 1\\
        1, \ \textbf{else}
    \end{cases}
\end{equation}
where, $\mathrm{\Xi}$ is a constant parameter. In this way, the activation threshold is adapted during dynamic conditions after every first spike using (14). Although this coding method can control the precision of the transmitted information, it also denotes that significantly more number of spikes needs to be transmitted for the same information that can be characterized by one spike. Hence, the selection criteria behind different coding methods build substantially on a trade-off between precision and the total number of spikes.
{
\subsection{Deployment Specifics}
As illustrated in Fig. \ref{FIG_12}(b), \texttt{Spike Talk} operates and learns by utilizing sparse output spike data to label remote features of neighboring DERs. These spikes are generated based on the neural coding schemes described previously, and can be locally deployed as parallel workloads in the controller of each DER. Although the selectivity of qualitative binary spike-based data is achieved post-coding, it is still derived from fixed-time sampled measurements equipped on each DER. For each event in the network (see Fig. \ref{FIG_12}(a)), the input spikes $\mathrm{ \Omega_i}$ are captured using the following coding strategy: 
\begin{eqnarray} \mathrm{ \Omega_i} = \begin{cases} 1 \ \textbf{if} \ \mathrm{\dot{I}_{k}(t)} \geq \mathrm{\kappa_2 e_k(t)} \\ 0 \ \textbf{else} \end{cases} 
\end{eqnarray} 
where, $\mathrm{e_i(t)} = [\mathrm{I_{ref}} - \mathrm{I_k(t)}]$ represents the current controller error as depicted in Fig. \ref{FIG_12}(a), and $\kappa_2$ is a tuning parameter. Given that the time constant of the current controller is lower than any other control stage in Fig. \ref{FIG_12}(a), the state-dependent threshold in (15) justifies the collection of high-resolution input data. This process is highlighted in Fig. \ref{FIG_12}(b), which is then further refined using various network-based thresholds and DER output $\mathrm{RC}$ dynamics to filter \textit{semantic} output data. The mechanism for online learning and continuous weight adaptation of each DER will be detailed in the next section.
}
\section{{Continual Online} Learning}
As already outlined in \cite{neuro}, the learning process in neuroscience is closely related to synaptic plasticity, which can then be attributed to the update of tie-line resistances as per the analogy in Fig. \ref{FIG_n}(d). 

\textit{\textbf{Remark V:} From a computational perspective, the notion of plasticity refers to either an increase or decrease of synaptic weights, however from a power network with a fixed topology, the tie-line resistance can instead be updated by adapting the dynamic droop value $\mathrm{\Delta R}$ in Fig. \ref{FIG_DER} for each DER.}

In comparison to supervised learning that is primarily steered on minimizing the error using backpropagation, Hebbian-based unsupervised learning operates on the relative firing time regime between the pre- and post-synaptic neurons. Going back to the context of DERs in Remark V, whenever there is an increase in the system demand, it increases the power generation from DERs, which is also denoted as an \textit{excitatory} response. Otherwise, it is called as an \textit{inhibitory} response corresponding to the decrease in power generation from each DER. Based on their definition and neuronal activity in computational neuroscience, \textit{excitatory} and \textit{inhibitory} responses correspond to the increase and decrease in adaptive droop $\mathrm{\Delta R}$, respectively.
\subsection{Voltage Regulation}
Voltage regulation as a secondary control objective is widely studied for microgrids and active distribution systems, which is performed differently for centralized or distributed communication topologies \cite{sahoo}. The basic idea behind this objective is to find a corresponding setpoint that allows equivalence of the mean of the DER voltages across the network close to its global reference $\mathrm{V_{ref}}$. Mathematically, it can also be denoted as $\mathrm{\sum_{i=1}^N}$ $\mathrm{V_i} \approx \mathrm{V_{ref}}$, with $\mathrm{N}$ indicating the total number of DERs.
Using Remark V, we can convert the nodal voltage dynamics into spikes via rate coding using (11). This can be simplified by analyzing the online learning framework in Fig. \ref{FIG_12}(c) performed at DER 1 in Fig. \ref{FIG_12}(a). 

Assuming $\mathrm{I_4}$ = 0, we can approximate the nodal voltages at DERs 2 \& 3 in Fig. \ref{FIG_12}(a) using the network equations and some approximations based on the global voltage regulation objective via:
\begin{equation}
    \begin{cases}
        \mathrm{V'_2(t) = V_1(t) + \psi_{12} r_dI_1(t)}\\
        \mathrm{V'_3(t) = V_1(t) + \psi_{13} (r_a + r_c)I_1(t)}
    \end{cases}
\end{equation}
where, $\mathrm{\psi_{12}}$ and $\mathrm{\psi_{13}}$ denote the regulation coefficient such that the average of all steady-state voltages of the network sum up to $\mathrm{V_{ref}}$. Using (16) as the threshold values for triggering spikes for each DER, we can firstly compute the \textit{voltage regulation spikes} $\mathrm{S_v(t)}$ with the help of the rate coding based triggering criteria in (11). Furthermore, it is worth notifying that the sum of all the regulation coefficients in a network must be equal to 1. As shown in Fig. \ref{FIG_12}(b), different static thresholds $\mathrm{V'_2}$ and $\mathrm{V'_3}$ are used to formalize the individual spike profile for DER 2 \& 3, which then act as an input for the calculation of adaptive droop $\mathrm{\Delta R}$.
\subsection{Power Management}
We will now consider different power management schemes, that are particularly driven by either generation/system cost \cite{cost}, unreliability (electrical and thermal) \cite{giampa}, etc. Since these conditions are dynamic, the energy management scheme augments a certain proportionality value $\mathrm{\alpha_k}$ such that the currents from the $\mathrm{k^{th}}$ DER is shared either equally or disproportionately among other DERs. Due to the low-pass filter disparity accounted into the measurements used in primary controller, there will be a delay in the rise/fall of currents in response to a given disturbance. Hence using Remark IV, we can use latency coding in (12) to investigate the competing rising time between currents with an activation voltage set up for each DER, given by:
\begin{eqnarray}
    \begin{cases}
        \mathrm{V'_{2ref}(t) = V_{ref} - R^2_d I_1(t)}\\
        \mathrm{V'_{3ref}(t) = V_{ref} - R^3_d I_1(t)}
    \end{cases}
\end{eqnarray}
where, $\mathrm{R^k_{d}}$ = $\mathrm{\alpha_k R_d}$. Using latency coding, we finally obtain the \textit{current sharing spikes} $\mathrm{S_i(t)}$, as shown in Fig. \ref{FIG_12}(c). It can be observed that there is a relative firing time difference $\mathrm{\Delta t = t_{S_v} - t_{S_i}}$ between the $\mathrm{S_v(t)}$ and $\mathrm{S_i(t)}$ layer, that ultimately determines the synaptic weight between the (pre-synaptic) DER and its (post-synaptic) neighboring DERs. Following the Hebbian unsupervised learning rule, the change in synaptic weight $\mathrm{\Delta w}$ for $\mathrm{k^{th}}$ DER varies as per:
\begin{eqnarray}
    \mathrm{\Delta w_k} =    
    \begin{cases}
        \mathrm{\sum_{f=1}^{F} \sum_{u=1}^{U} A_+ e^{\frac{\Delta t}{\tau_k}}}, \ \textbf{if} \ \mathrm{\Delta t > 0}\\
       \mathrm{\sum_{f=1}^{F} \sum_{u=1}^{U} A_- e^{\frac{\Delta t}{\tau_k}}}, \ \textbf{else}.
    \end{cases}
\end{eqnarray}
where, the magnitudes $\mathrm{A_+}$ and $\mathrm{A_-}$ depend on the current synaptic weight. Moreover, $\mathrm{f}$ = \{1, 2, 3,...$F$\}, $\mathrm{u}$ = \{1, 2, 3,...$L$\} represent the labels of pre-synaptic and post-synaptic spike times. Furthermore, $\mathrm{\tau_k}$ determine the $\mathrm{RC}$ time constant of the $\mathrm{k^{th}}$ DER. The initial values of $\mathrm{R}$ in $\mathrm{\tau}$ can be obtained and updated using the optimal solution of (9). In this way, the relative firing time difference $\mathrm{\Delta t}$ is exploited with respect to each DER time constant for inferring the global information for steering energy-efficient data-driven control of microgrids. For incremental deployment of the Hebbian learning policy into microgrids using (18), we modify the primary controller droop using:
\begin{eqnarray}
\mathrm{\undb{V}_{ref} = \undb{V}(t) - \undb{I}(t)[\undb{R}_d - \undb{a}\Delta \undb{w}(t)]}
\end{eqnarray}
where, $\undb{a}$ is a system matrix of a normalization variable. Using (18)-(19), we steer an on-device grid-edge controller for each DER, that drives using the nodal and network equations in (16)-(17) to update their respective weights using the spike timing dependent plasticity (STDP) \cite{neuro}. Its system-wide implementation is governed using the power flow dynamics seen by each DER, which is the crux behind \textbf{\texttt{Spike Talk}}.
\subsection{Theoretical Guarantees}
Due to the simple nature of Hebbian learning, only the firing time differences between the pre- and post-synaptic activity may not provide a substantial justification on how the learning process leads to be effective \cite{neft}. However, we exploit the microgrid dynamics and the statistical properties of circuit theory in the considered system in correlation with the neuronal circuits. Since Hebbian plasticity exploits these statistical features and domain knowledge in a certain way, it can be categorized as unsupervised learning.

Let us consider a single rate LIF neuron input model in (9), where the synaptic current $\mathrm{I(t)}$ is the output $\mathrm{y}$, with the membrane potential $\mathrm{V_{mem}(t)}$ of multiple neurons acting as inputs $\mathrm{x_1(t),...,x_N(t)}$. Using (9), we can express the output using a response function $\mathrm{f}$, given by:
\begin{eqnarray}
    \mathrm{y = f(\sum_{i=1}^N w_i x_i)}.
\end{eqnarray}
Since we ascribe Hebbian plasticity in terms of the evolution of the synaptic weight updates, we obtain its rate of change using (19):
\begin{eqnarray}
    \mathrm{\frac{dw_i}{dt} = \eta x_i y = \eta x_i \sum_{j=1}^{N} w_j x_j}. 
\end{eqnarray}
Considering the vector form of (20) and training over an average $\langle...\rangle$ over discrete training set, we now use the analogy chart in Fig. \ref{FIG_n}(d) and Remark V to obtain $\Delta \undb{R}$ using the weight update policy: 
\begin{eqnarray}
  \mathrm{ \frac{d\undb{w}}{dt} = \langle\eta \undb{V} \undb{V}^T \undb{Y}\rangle = \eta \langle\undb{V}\undb{V}^T\rangle\undb{Y} = \eta C \undb{Y}}.
\end{eqnarray}
where, $\undb{Y}$ is the system admittance matrix with tie-line resistances $\mathrm{r_{kl}}$. Furthermore, $\mathrm{C}$ = $\langle\undb{V}\undb{V}^T\rangle$ denote the correlation matrix. As a result, a system with $\mathrm{N}$ DERs can be simply modeled as $\mathrm{N}$ coupled linear differential equations. Apart from the stable network models such as Oja's rule \cite{oja} or generalized Hebbian algorithm \cite{hebb}, for $\mathrm{t} \rightarrow \infty$, the solution for synaptic weights can be given by:
\begin{eqnarray}
\mathrm{\undb{w}(t) \approx e^{\eta \alpha^* t} \undb{c}^*}
\end{eqnarray}
where, $\alpha^*$ and $\undb{c}^*$ denote the largest eigenvalue and eigenvector of $\mathrm{C}$, that corresponds to computing the first principal component of the inputs. It should be noted that $\eta$ can act as a normalization parameter using the total capacity of the system $\mathrm{P_{t}}$. Finally, basic circuit theory performed on the considered system in (22) provides theoretical guarantees of a stable solution in (23), which can be extended to perform a full principal component analysis (PCA) by adding more post-synaptic neurons. In this way, the statistical aspects of the input spikes and the convergence of the output signals using \textbf{\texttt{Spike Talk}} can be justified as a networked unsupervised plasticity-based learning with a significant energy-efficient and ultra-low latency footprint.
\section{Performance Evaluation}
\begin{figure}[!thb]\centering
	\includegraphics[width=\linewidth]{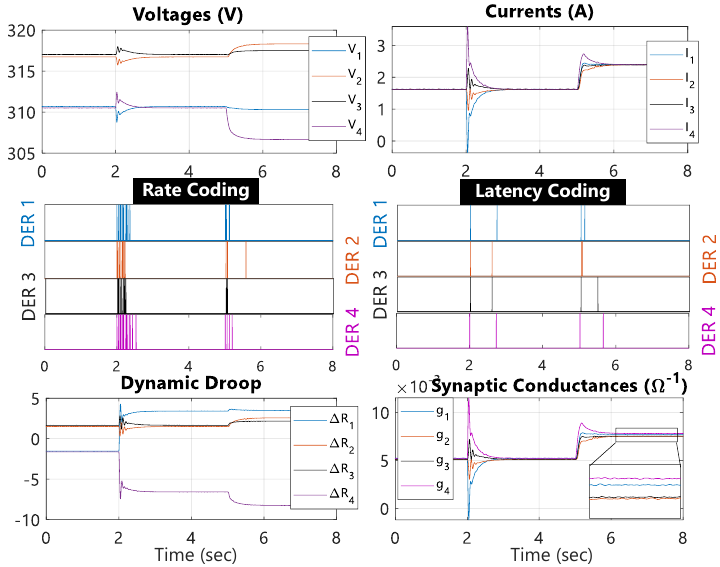}
	\caption{Case I -- Performance validation of \textit{Remark IV} on our hypothesis behind rate and latency coding for processing of spikes corresponding to voltage regulation and current sharing, respectively.}
    \label{FIG_c1}
\end{figure}
\begin{figure*}[!thb]\centering
	\includegraphics[width=0.88\linewidth]{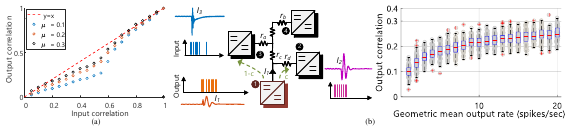}
	\caption{(a) Sensitivity of spikes generation in the output of a DER due to disturbances from the input comprising of the intermittent renewable energy sources -- a higher correlation factor $\mu$ is adjudged to be providing the same spike density in the output, {(b) Correlation detection to determine the correlation of output spikes between DER 1 and 2}.}
    \label{FIG_c12}
\end{figure*}
In this section, we analyze and evaluate the performance of \texttt{\textbf{Spike Talk}} in the system shown in Fig. \ref{FIG_DER}(b) comprising of $\mathrm{N}$ = 4 DERs. The neuro-edge computing tool designed using (19) is integrated into each DER to analyze performance variations under different neural coding schemes. Four test cases were considered, where the modeling of the system in \ref{FIG_DER}(b), generation of spikes using (11)-(14), Hebbian learning using (16)-(19) and the entire framework was carried out in MATLAB/Simulink environment. We consider global voltage regulation as well as proportionate current sharing as the secondary controller objectives for each DER. With these objectives federating the spike generation as well as weight updates, we execute the online learning framework in Fig. \ref{FIG_12}. The system and control parameters can be found in Appendix.

In Case I, we investigate the performance of Spike Talk in Fig. \ref{FIG_c1} to test our hypothesis on end-to-end learning using spikes derived from latency coding, whereas edge learning from rate coding. Exploiting the modeling analogy between neuronal architecture and DERs in Fig. \ref{FIG_n}, we leverage rate coding scheme in (11) with physics-informed firing and reset thresholds in (16). As shown in Fig. \ref{FIG_c1}, the rate coded spikes achieve global voltage regulation with an average around 315 V. These spikes are commonly referred to as the \textit{voltage regulation spikes} in Fig. \ref{FIG_12}(c). Since the dynamic droop is accountable for both the objectives, we then formulate proportionate current sharing using (13) as the time-trigger for each DER currents computed locally. Using (13), we exploit the latency of the second spike from the first spike, that basically corresponds to the calculation of the oscillatory frequency for a given disturbance due to heterogeneous control parameters. Hence, we incorporate a pseudo-threshold for each DER, $\mathrm{V'_{2ref}}$ and $\mathrm{V'_{3ref}}$ in (17), which assists in determining the relative performance of the neighboring DERs and their response times. Since heterogeneous droop parameters imparts different overshoot and rise time to each DER, we curate latency coding to determine these performance metrics. This argument can be clearly verified with the latency coding scheme in Fig. \ref{FIG_12}(b) where different peaks resulted into different temporal positions of the second spikes. It is worth notifying that the first spike is always triggered to any disturbance, which then translates as $t_0$ for each event. Finally using the \textit{voltage regulation spikes} and \textit{current sharing spikes}, we compute the weight adaptation $\mathrm{\Delta w_k}$ using (18). After customizing $\Delta \undb{w}$ into $\Delta \undb{R}$ using (19), we plot the dynamic droop in Fig. \ref{FIG_c1} that adapts itself based on the disturbance. By visualizing the synaptic conductance of each DER using (5), there is a steady-state difference between them, which is zoomed in Fig. \ref{FIG_c1} that not only validates our hypothesis in Remark IV, but also validates the modeling analogy between neurons and DER. The steady-state difference between the synaptic conductance can be accounted to the resistive drop in the tie-lines, which consequently demands slightly more current than the overall demand.

Since most event-driven techniques are primarily designed based on the output disturbances for each DER due to the output RC circuit in Fig. \ref{FIG_n}(b), we introduce a transient disturbance in the input of DER 1 at t = 2 sec in Fig. \ref{FIG_c1} to determine a correlation factor $\mu$ of the system response to input events. Defining the input and output correlation rates $\mathrm{\rho_i}$ and $\mathrm{\rho_o}$ respectively as the average rate of spikes over a given time interval, the correlation factor between the input and output of $\mathrm{k^{th}}$ DER can be split using:
\begin{eqnarray}
    \mathrm{\kappa^{-1}\dot{V}_k[t_h] = \mu \rho_iI_{in}[t_h] - \rho_o I_{o}[t_h].}
\end{eqnarray}
With $\mathrm{t_h}$ being the firing instant in (24), we leverage the voltage dynamics of each DER, which is closely related to the input as well as output current of a DER to approximate the co-relationship of spikes. By mapping different correlation rates against each other, we can demonstrate almost a quasi-linear relationship between both ends in Fig. \ref{FIG_c12}(a). It can be seen that the output correlation rate diverges to the lower end away from the linear plot for $\mathrm{\rho_i}$ within [0.4, 0.7]. This is attributed to the fact that the spike generation gets compromised by the discontinuous mode of operation, which is not aligned with the assigned firing threshold and vary based on the loading condition. 

{Although online weight adaptation in each DER does not rely on estimating the currents from remote sources, it is still vital to assess the proportion of currents from remote DERs. This aspect can be simply answered by another neuroscience phenomena, known as \textit{coincidence detection} \cite{nat}. This basically means that the spike train correlations between multiple neurons can be quantified by statistical analysis of the spike trains and their correlation coefficients. For example, $\mathrm{I_1}$ in Fig. \ref{FIG_c12}(b) is composed of different proportions of $\mathrm{I_2}$, $\mathrm{I_3}$ and $\mathrm{I_4}$, which is impossible to predict without having sufficient global information. As a result, $c$ in Fig. \ref{FIG_c12}(b) is unknown. Since (16) anyway require the regulation coefficients to ensure global voltage regulation within a specified limit, these regulation co-efficients for DER 1 can be determined using:
\begin{eqnarray}
    \mathrm{I_1} = \mu_1 + \sigma_1 \underbrace{[(1-c)\Psi_k(t) + c\Psi_{co}(t)]}_{\psi_{1k}}
\end{eqnarray}
where, $\mathrm{\mu_1}$ is the temporal average of current $\mathrm{I_1}$ with the next term representing the variance due to Gaussian fluctuations, that comprises of two weighted factors: $\mathrm{\Psi_k}$ which is independent for the remote $\mathrm{k^{th}}$ DERs, and $\mathrm{\Psi_{co}}$ which is common for all the DERs. Furthermore, the input correlation coefficient (0 $\mathrm{\leq c \leq}$ 1) set the relative weight of the proportion of currents from remote DERs, whereas $\sigma_1$ is the normalized variance of the total input current. These parameters have been normalized in such a manner that the sum of regulation coefficients naturally is 1, owing to the common variance term in (25) for current from each DER. A detailed proof and analysis behind computing these variables using the variability and co-variability of membrane potentials and spike trains will be covered in the future scope of this work.

Since coincidence detection is highly dependent on the rate of spikes, we illustrate a case study on correlation-rate relationship in Fig. \ref{FIG_c12}(b). In the case study, a high probability of spikes is seen with an increase in the geometric mean rate of spikes, given by $\mathrm{\sqrt{v_1 v_k}}$, where $\mathrm{v_k}$ suggest the average rate of output spikes from $\mathrm{k^{th}}$ DER. With the correlation ranging from 0 to 0.3, many perturbation techniques \cite{nat} can be used to invoke the rate of spikes from each DER without compromising system performance.
}

\begin{figure*}[!thb]\centering
	\includegraphics[width=0.8\linewidth]{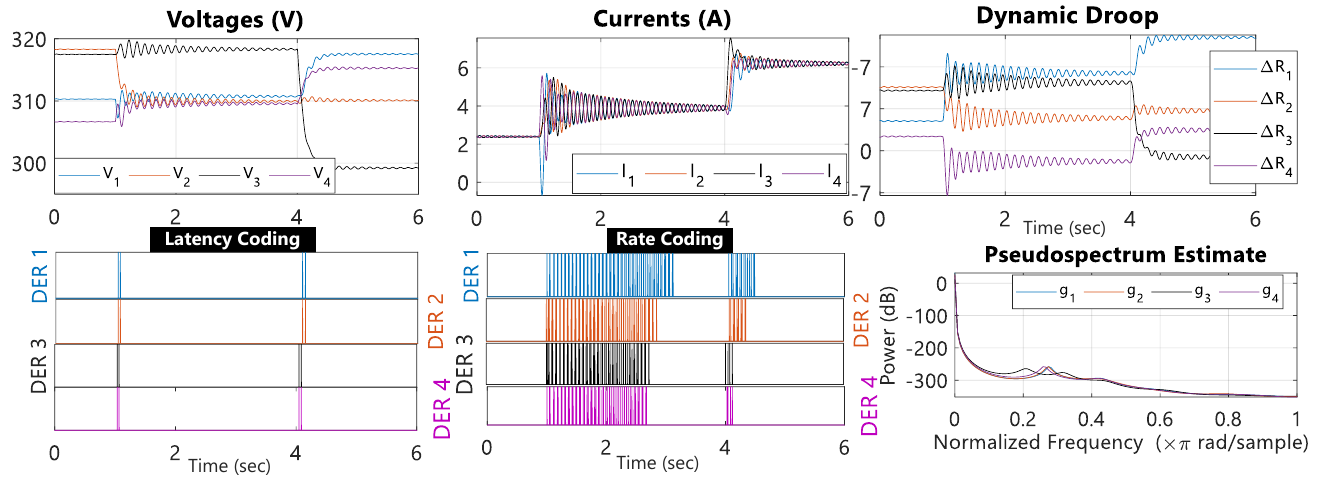}
	\caption{Case II -- Swapping the neural coding schemes to evaluate the system performance results into instability. The pseudospectrum estimate of the synaptic conductance $\mathrm{g_k}$ for each DER depicts the strength of the oscillating frequency, which is a byproduct of mistimed temporal spike patterns.}
    \label{FIG_c2}
\end{figure*}

In Case II, we swap the coding scheme settings previously used in Case I to evaluate the considered system's performance in Fig. \ref{FIG_c2}. As a result, we use latency coding for global voltage regulation and rate coding for equal current sharing. As it can be seen in Fig. \ref{FIG_c2}, although the coding schemes in itself perform normally, their decoding into the dynamic droop $\Delta \undb{R}$ using (18) exhibit an oscillatory component, which potentially makes the system operation unstable. This is quite evident from the voltage and current responses for all the DERs, which affirms our hypothesis of the coding assignment for different secondary control objectives in Remark IV. Upon plotting a pseudospectrum estimate of the synaptic conductance $\mathrm{g_k}$ in Fig. \ref{FIG_c2}, we deduce the strength of the oscillating frequency emanating from relatively mistimed spike positions. The mistimed notion of spikes can be interpreted as $\mathrm{\Delta \undb{R}(t+\tau_d)}$, where $\mathrm{\tau_d}$ represent the synaptic delay that can range from 50-150 ms \cite{lilli}. As a result of this delay, the interactions between voltage control loops at a system level increases, that results into an unstable condition. Since it directly affects the ultra-low latency capability, the suitability of the assigned coding schemes is affected for both control objectives in Remark IV to instill a satisfactory system performance.

\begin{figure}[!thb]\centering
	\includegraphics[width=\linewidth]{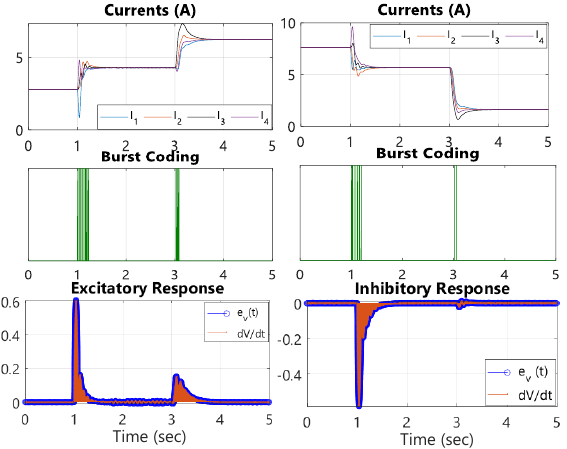}
	\caption{Case III -- Performance evaluation of burst coding in \texttt{\textbf{Spike Talk}} using (14). Although burst coding has a higher degree of biological plausibility, it always reflects on high frequency events, that does not necessarily cater to all the disturbances. Hence, its applicability is limited to high frequency and short inter-spike interval (ISI) events.}
    \label{FIG_c3}
\end{figure}
\begin{figure}[!thb]\centering
	\includegraphics[width=\linewidth]{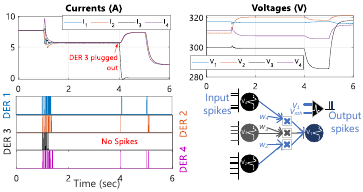}
	\caption{Case IV -- Plug-and-play capability using \texttt{\textbf{Spike Talk}}, when DER 3 is plugged out of the system at t = 4 sec. Since the network topology is modified with outage of DER 3, no spikes are recorded since the neuro-edge computational unit at DER 3 switches to an inert stage with $\mathrm{V_3}$ and $\mathrm{I_3}$ forced to zero. From a system-level computational perspective, the plugged out DER clears away itself from the processing of the output.}
    \label{FIG_c4}
\end{figure}
As elaborated in the last section, burst coding relies on high frequency events using (14), which inherits an incremental pattern in the signal magnitude in either positive or negative direction over consecutive time intervals. As we deploy each DER with a neuro-edge computational tool modeled as a decentralized secondary controller, high frequency events in power systems do not cater to a response from a secondary controller, which has a typical settling time response in the order of seconds \cite{sahoo2017distributed}. As a result, the applicability of burst coding is quite limited despite its biological accuracy and reliability of information transfer. However due to its multi-fold advantages with sparse data and ingrained energy-efficient computational features, we programmed burst coding into the considered system to evaluate system performance in Fig. \ref{FIG_c3}. As it can be seen in Fig. \ref{FIG_c3}, burst coding instills a satisfactory current sharing performance for an increase as well as decrease in load in Case III. It can be argued using the Hebbian plasticity rule in Remark V for DER 1 that an increase in system demand causes an \textit{excitatory} response with the error excitation in the positive region, which is evident from Fig. \ref{FIG_c3}. Similarly for decrease in load, an \textit{inhibitory} response can be observed with the voltage error excitation in the negative region at all times. Furthermore, the voltage error is always bounded by the trajectories of $\mathrm{\dot{V}_1}$ using (11).

In Case IV, the plug-and-play capability of Spike Talk is investigated in Fig. \ref{FIG_c4}. It can be seen that when DER 3 is plugged out of the considered system at t = 4 sec, the rest of the DERs increase their share of current sharing among each other. For a decrease in load at t = 5 sec, the current sharing is still intact with DER 3 as the inert neuron. Using our previous argument in Remark III, it can then be deduced that the network level computational overhead is significantly decreased due to the modeling analogy in Fig. 3(d). It is shown in Fig. \ref{FIG_c4} that outage of DER 3 is efficiently handled by the Spike Talk architecture, which automatically excludes $\mathrm{V_3 w_3}$ from (9) by initializing $\mathrm{I_3}$ = 0. From a computation perspective on a system level, restricted power flow leading to absence of spikes from DER 3 also implies a dropout of computation to be carried out in DER 1. In this way, we ensure a highly accurate, sparse as well as low computational overhead for coordination of DERs only using digital spikes emanating out of real-valued local analog measurements.
\section{Conclusions and Future Scope of Work}
We unravel a novel data-driven co-transfer technology \texttt{\textbf{Spike Talk}} that exploits modality of power and information for sparse data-driven coordinated control of microgrids. This co-transfer principle is inspired from computational neuroscience, where we decipher a modeling analogy between the elements of a brain and power systems. Using this structural complementary mapping, we design a decentralized secondary controller for each DER, that is federated online by power flow in the network. As a result, this completely dismisses the cyber layer and breaks the traditional hierarchy of cyber-physical microgrids to wipe out all the cyber reliability issues and vulnerabilities. In simple words, the dynamic droop parameter in the primary controller of each DER is adapted by Hebbian unsupervised learning based plasticity rules, where localized input data is selectively collected and processed using the neural coding schemes. These coding schemes introduce a sparse data analysis stage that converts synchronous real-valued measurements into asynchronous digital spikes. With increased investments in re-conductoring and smart conductors \cite{forbes}, \texttt{\textbf{Spike Talk}} will be an enormous grid-edge asset that unlocks power grid's potential in the field of energy efficiency, digitalization and electrification without any compromise in system reliability and performance. As we establish the selection of neuronal weights and thresholds based on each DER's dynamics and maximum capacity respectively, it becomes fairly simple to design and can be incrementally integrated into the controller of each DER.

As a future scope of work, the remaining operational elements of \texttt{\textbf{Spike Talk}}, such as stability, multiplexing of information along tie-lines  will be revealed through extensive validation and theoretical guarantees. 

\section*{Appendix}
It should be noted that the system and control parameters are consistent for each DER, unless stated otherwise.

$\undb{V}_{ref}$ = 315 V, $\mathrm{C_1}$ = 450 $\mu$F, $\mathrm{C_2}$ = 500 $\mu$F, $\mathrm{C_3}$ = 480 $\mu$F, $\mathrm{C_4}$ = 520 $\mu$F, $\mathrm{r_a}$ = 0.5 $\Omega$, $\mathrm{r_b}$ = 0.25 $\Omega$, $\mathrm{r_c}$ = 0.6 $\Omega$, $\mathrm{r_d}$ = 0.8 $\Omega$, $\undb{I}_{max}$ = 10 A, $\Delta \undb{V}$ = 20 V, $\undb{R}_d$ = 2, $\mathrm{f_{sw}}$ = 20 kHz, $\undb{a}$ = 2, $\Xi$ = 4.8, $\psi = \begin{bmatrix}
    0 & 0.6 & 0.4 & 0\\
    0.3 & 0 & 0.4 & 0.3\\
    0 & 0.4 & 0.25 &\ 0.35\\
    0 & 0.15 & 0.25 & 0.6
\end{bmatrix}$, $\kappa_1$ = 0.9, $\kappa_1$ = 0.75.





\ifCLASSOPTIONcaptionsoff
  \newpage
\fi



%

\bibliography{Reference.bib}
\bibliographystyle{IEEEtran}
%








\end{document}